\documentclass{article}
\usepackage{smc2021}
\usepackage{multirow}
\usepackage{paralist}
\usepackage[figure,table]{hypcap}
\usepackage{graphics}

\usepackage[whole]{bxcjkjatype}

\usepackage{alphabeta}
\usepackage{arabtex}
\usepackage[LFE,LAE,LGR,T2A,T1]{fontenc}
\usepackage[greek, russian, main=english]{babel}

\usepackage{subcaption}
\usepackage{xcolor}
\def\papertitle{Tracing Back Music Emotion Predictions to
Sound Sources \\ and Intuitive Perceptual Qualities}

\author[1]{\mbox{\firstname{Shreyan}\lastname{Chowdhury}\email{shreyan.chowdhury@jku.at}}}
\author[1]{\mbox{\firstname{Verena}\lastname{Praher}\orcid{0000-0001-5466-7829}}}
\author[1,2]{\mbox{\firstname{Gerhard}\lastname{Widmer}\orcid{0000-0003-3531-1282}}}

\affil[1]{\department{Institute of Computational Perception}\institution{Johannes Kepler University}\city{Linz}\country{Austria}\affiliationtype{University}}
\affil[2]{\department{LIT AI Lab}\institution{Linz Institute of Technology}\city{Linz}\country{Austria}\affiliationtype{University}}

\completesetup

\title{\papertitle}
\begin{document}
\capstartfalse
\maketitle
\capstarttrue
%


\begin{abstract}
Music emotion recognition is an important task in MIR (Music Information Retrieval) research. Owing to factors like the subjective nature of the task and the variation of emotional cues between musical genres, there are still significant challenges in developing reliable and generalizable models. One important step towards better models would be to understand what a model is actually learning from the data and how the prediction for a particular input is made. In previous work, we have shown how to derive explanations of model predictions in terms of spectrogram image segments that connect to the high-level emotion prediction via a layer of easily interpretable perceptual features. However, that scheme lacks intuitive musical comprehensibility at the spectrogram level. In the present work, we bridge this gap by merging audioLIME -- a source-separation based explainer -- with mid-level perceptual features, thus forming an intuitive connection chain between the input audio and the output emotion predictions. We demonstrate the usefulness of this method by applying it to debug a biased emotion prediction model.
\end{abstract}

\section{Introduction}\label{sec:introduction}




The quest for interpreting the inner workings of ``black-box'' models and explaining their predictions has led to many recent advances in the area of explainable AI and is becoming an increasingly important staple of all AI subfields. 
Not only do model explanations help in enhancing trust in the model in applications where its predictions are critical decisions like creditworthiness or medical diagnosis, but they can also often reveal telling signs of algorithmic bias ~\cite{Rieger2020Penalizing, Samek2019Biases}. While algorithmic decisions in the field of MIR are not as life-critical as medical diagnosis, in today's era of music streaming and recommendations, they can have far-reaching effects on diverse audiences, creators, and artists alike.
For instance, machine-learning-based music recommender systems have been shown to exhibit severe biases towards or against certain user groups ~\cite{ekstrand2018BiasRecSys}. Interpretable explanations of these models or their predictions would be extremely helpful for identifying such biases.

Previous work on interpretability in MIR has dealt with tasks such as music tagging using self-attention \cite{won2019toward} and transcription using invertible neural networks \cite{kelz2019towards}, and
post-hoc explanations for music content analysis have been used to understand what a genre classifier~\cite{choi2016explaining} or a singing voice detector~\cite{mishra2017local, mishra2020reliable, mishra2018a,  mishra2019GANs, mishra2018eusipco}
have learnt. 
More recently, audioLIME has been proposed 
\cite{Haunschmid2020MML, ribeiro2016should} and has shown promise in explaining tagging models \cite{Haunschmid2020arxiv} as well as recommendation models \cite{Melchiorre2020}. 


However, explanations of \textit{music emotion recognition} systems have received relatively less attention notwithstanding the importance of this task in the areas of musical analysis and recommendation. \cite{DeBerardinis2020} used models trained with different combinations of sound sources to deduce the importance of each source to emotion predictions. We previously proposed using an intermediate layer of \textit{mid-level perceptual features} \cite{Aljanaki2018Midlevel} to explain music emotion recognition models through a linear connection between the intermediate and final layers  \cite{Chowdhury2019}. We followed this up with a two-step explanation approach \cite{Haunschmid2019TwoLevel} to further explain the predictions in the mid-level layer using components from the input spectrogram. This two-level scheme used LIME (\textit{Local Interpretable Model-agnostic Explanations})~\cite{ribeiro2016should} 
to construct explanations for the mid-level predictions in terms of specific patches of the input spectrogram (which were obtained via image segmentation).
While this gave us the regions in the spectrogram contributing most to a particular prediction, these regions did not hold any musical meaning by themselves. As a result, it is difficult to comprehend these explanations in terms of meaningful or intuitive concepts.

In this work, we bridge this gap by merging the audioLIME method, which uses sound sources as explanatory features, with the approach of mid-level features, to obtain comprehensible explanations from the input audio as well as from the perceptual layer. It thus forms an intuitive connection of hierarchical explanations from low-level constituent sources of audio to the high-level emotion predictions through the intermediary mid-level layer, all of which have a musical interpretation. 


We believe that explainability is particularly important for developing better music emotion recognition algorithms since it is often difficult to identify misclassifications and biases in this task because of its inherent subjectivity and inter-annotator variability. As an example of real-life application of our method in understanding the potential cause of bias in an emotion model, we demonstrate how a model that has seen few examples of a genre during training results in a pattern of errors on the test set that contains examples of this genre. We trace this pattern of errors back to a particular mid-level feature and this feature to a particular source in the input. Doing so allows us to predict how the model would change when retrained with a balanced training set. We can then qualitatively verify that the retrained model has in fact changed in the way that we expected from our explanations.

\section{Two-level Explanations}

Our proposed two-level system will explain emotion predictions by first tracing them back to the most relevant mid-level features and in a second step explain the intermediate mid-level layer via audio sources. We will first describe each of the parts separately from the lowest level (audio sources) to the highest level (emotion predictions) and then put together all the parts in Section~\ref{sec:put_together}.

\subsection{Explaining via Audio Sources: audioLIME}

In order to explain individual mid-level  predictions we make use of audioLIME, a recently introduced approach based on LIME~\cite{ribeiro2016should} for interpreting models in MIR~\cite{Haunschmid2020arxiv, Haunschmid2020MML}. LIME uses simplified inputs based on a set of human interpretable features (depending on the domain of the task -- e.g., superpixels for images) to train a simpler explanation model $g$ in order to explain a more complex, potentially deep model $f$. Previous approaches based on LIME have used time-, frequency-, or time-frequency segments~\cite{mishra2017local}, or segments computed by an image segmentation algorithm~\cite{Haunschmid2019TwoLevel}. audioLIME introduced a new type of interpretable features: sound sources estimated by a music source separation algorithm. In other words, the audioLIME explanation for a given prediction will be in the form of particular sound sources (and possibly specific temporal segments -- which we do not use right now), telling us that it is some sonic aspects of these sources that seem to be influential. 
In our case, the source separator is the pretrained music source separator spleeter~\cite{spleeter2020}, thus the explanatory sound sources will be (what the source separator believes are) individual instruments.

\subsection{Explaining via Mid-level Perceptual Features}\label{sec:midlevel_exp}
\textit{Mid-level features} are perceptual qualities or descriptors that emerge from low-level musical building blocks such as timbre, beat structure, harmony, etc. They are more subjective than the low-level features, thus difficult to model using hand-crafted feature extractors, but musically discernible enough to have many people reach a high agreement in annotations. Examples include qualities such as perceived melodiousness or rhythmic complexity \cite{Aljanaki2018Midlevel}. They, therefore, form a suitable choice for an intermediary to higher-level concepts like emotion.

This idea was first used in \cite{Chowdhury2019} to explain emotion predictions. We use a similar approach here, but we use the more recently introduced receptive-field regularized ResNets \cite{Koutini2019Mediaeval, Koutini2020arxiv} to model the emotions and the mid-level features. In addition, we learn the emotions and mid-level features from two separate datasets using multi-task learning, i.e., jointly learn to predict mid-level features and high-level emotions, allowing us to be flexible with our train and test domains.

The mid-level layer is the penultimate layer of the model, with a linear transformation between it and the final (emotion) layer, thus making the connection interpretable. It is learnt end-to-end from audio spectrograms by optimizing on the combined loss from the emotion and mid-level layers.

Emotion predictions can be interpreted by looking at \textit{effects plots}, which are visual representations of the contribution of each mid-level feature to the final emotion prediction, calculated as the product of the feature value and the weight joining it to the emotion. These indicate the relative influence of the individual features on the final prediction.

\subsection{Putting it All Together: Intuitive Two-level Explanations}
\label{sec:put_together}

For a given prediction, we then work backwards. First, we obtain the mid-level explanation of an emotion by computing the effects. The larger the effect of a mid-level feature, the larger is the contribution of that feature to the emotion prediction. Next, we compute the audioLIME explanations for the mid-level feature with the largest effect (we can in principle compute audioLIME explanations for all features to obtain a more diverse explanation, depending on the application). 
Given these two explanations, we can describe a prediction as being arrived at by the model due to the explanatory mid-level feature, which is in turn most influenced by the input component given by audioLIME.





\section{Experimental Setup}
For our analysis with emotion explanations, we first need to train the ``explainable'' models, which have the penultimate mid-level layer connecting linearly to the final emotion outputs. This section describes the datasets and the training procedure for such models.

\subsection{Datasets}
\label{sec:data}

For our experiments we are using three different datasets:

\subsubsection{Mid-level Perceptual Features Dataset}

The \textit{Mid-level Perceptual Features Dataset} introduced in~\cite{Aljanaki2018Midlevel} consists of 5000 song snippets with annotations between 1 and 10 for the mid-level descriptors \textit{melodiousness, articulation, rhythmic complexity, rhythmic stability, dissonance, tonal stability}, and \textit{modality} (called ``minorness'' here). We use this dataset to train the intermediate layer of our emotion model.

\subsubsection{DEAM: Database for Emotional Analysis in Music}

The \textit{DEAM} dataset~\cite{aljanaki2017developing} is a dataset of dynamic and static valence and arousal annotations. It contains 1,802 songs (58 full-length songs and 1,744 excerpts of 45 seconds) from a variety of Western popular music genres (rock, pop, electronic, country, jazz, etc). In our experiments, we use the static emotion annotations, which are continuous values between 0 and 10.

\subsubsection{PMEmo: Popular Music with Emotional Annotation}

The \textit{PMEmo} dataset~\cite{Zhang2018_PMEmo} consists of 794 chorus clips from three different well-known music charts. 
The songs were annotated by 457 annotators with valence and arousal annotations separately for dynamic and static.
In our experiments, we use static labels, which are continuous values between 0 and 1.

\subsection{Model Training}

\begin{table}[t]
\small
\resizebox{\columnwidth}{!}{%
  \begin{tabular}{l|l|l|l|l}
    \multirow{2}{*}{} &
      \multicolumn{2}{c|}{Arousal} &
      \multicolumn{2}{c}{Valence} \\
    
    & RMSE & R2 & RMSE & R2  \\
    \hline
    P(P - bl) & 0.23  & 0.61 & 0.25  & 0.41  \\
    P(P) & 0.25 $\pm$ 0.03 & 0.60 $\pm$ 0.10 & 0.31 $\pm$ 0.04 & 0.40 $\pm$ 0.14  \\
    
    D(P) & 0.27 $\pm$ 0.01 &  0.50 $\pm$ 0.03 & 0.33 $\pm$ 0.00 & 0.30 $\pm$ 0.02  \\
    D(D) & 0.26 $\pm$ 0.01 &  0.49 $\pm$ 0.02 & 0.22 $\pm$ 0.01 & 0.51 $\pm$ 0.04  \\
    
    D+P(P) & 0.23 $\pm$ 0.01 & 0.65 $\pm$ 0.02 & 0.28 $\pm$ 0.00 & 0.50 $\pm$ 0.02 \\
    D+P(D) & 0.26 $\pm$ 0.01 & 0.50 $\pm$ 0.03 & 0.23 $\pm$ 0.01 & 0.48 $\pm$ 0.02 \\
  \end{tabular}%
  }
  \caption{\small Emotion prediction performance with our ``explainable'' model trained and tested on difference datasets --  P: PMEmo, D: DEAM. The dataset inside the parentheses is the test dataset. The top row is the baseline performance from~\cite{DeBerardinis2020}.  
  \label{tab:emotion_results}}
\end{table}

The mid-level and emotion model is trained end-to-end using audio spectrograms as inputs and optimizing on the combined loss from the mid-level and emotion layers. The batch size is 16 and contains 8 samples from the Mid-level dataset and 8 samples from either the DEAM or the PMEmo dataset. The loss function is the mean squared error. The learning rate is $10^{-3}$ with cosine annealing, and we perform early stopping on a validation set as regularization. We use the Adam optimizer~\cite{kingma2014adam}.

The inputs are log-filtered spectrograms (149 bands) of 40-second audio clips peak normalized and sampled at 22.05 kHz with a window size of 2048 samples and a hop length of 704 samples, resulting in 149$\times$1252-sized tensors. If a clip is longer than 40 seconds, we take a random snippet, and if it is shorter, it is looped to 40 seconds.

The labels are scaled to the range $[-1, 1]$ for all three datasets. Therefore, an RMSE of 0.26 would represent 13\% error. We split the train and test sets such that they have mutually exclusive sets of artists. A summary of the emotion prediction performance can be found in Table~\ref{tab:emotion_results}.



\section{Evaluation of Explanations}

Essentially, there are three targets for empirical evaluation: the two individual components of our two-level explanation framework, and the final composite explanations produced by the model. Regarding the former, the higher level -- explanations of emotion predictions in terms of mid-level perceptual features -- has already been discussed at length in our previous paper~\cite{Chowdhury2019}. We showed how effects plots can give insight into the relative importance of various mid-level qualities. 
The lower level -- using audioLIME to explain mid-level feature predictions via audio sources -- is a new concept, and the experiments in the following section are intended to validate it.
Empirical evidence for the usefulness of the complete, two-level explanation model, finally, will be presented in the form of a study, in Section~\ref{sec:model_debugging} below, where we demonstrate how these explanations can help us debug a biased prediction model by gaining insight into what the sources of its problems are.

%
%

\subsection{Explaining Mid-level Features via Sound Sources}

Evaluating the quality of explanations is a hard task since there is no consensus on what makes a good explanation, with a variety of desired aims and properties proposed in literature ~\cite{bhatt2020evaluating, lipton2018, tintarev2012}). We build our evaluation of audioLIME explanations for the mid-level layer on two metrics, (a) \textit{fidelity} as proposed by Ribeiro et al. along with LIME~\cite{ribeiro2016should} and, (b) \textit{complexity}, a recently proposed metric for feature-based model explanations~\cite{bhatt2020evaluating}. 

Fidelity measures how well the local model $g$ (the explainer) approximates the global model $f$ (the model up to the mid-level layer in our case)~\cite{ribeiro2016should} and is computed using the coefficient of determination between the local and global model's predictions as in the original LIME implementation\footnote{\url{https://github.com/marcotcr/lime/}}. 

In addition to high fidelity, low complexity is desired. The most complex explanation would be the one where all $d$ features get the same attribution (i.e., all weights $g(f, x)_i$ of the linear explanation model $g$ are the same). The simplest explanation concentrates all attribution on one feature. To measure complexity, a probability distribution $P_g$ is defined:

\begin{equation}
P_g(i) = \frac{|g(f, x)_i|}{\sum_{j \in [d]}{g(f, x)_j}}
\end{equation}

Complexity is then defined as the entropy of this distribution~\cite{bhatt2020evaluating}. We compare the complexity per dataset with a random baseline, which is obtained by creating ``random'' explanations with feature weights drawn from a uniform distribution.

\begin{figure*}[!ht]
    \centering
    \begin{subfigure}{.5\textwidth}
        \centering
        \includegraphics[width=\columnwidth]{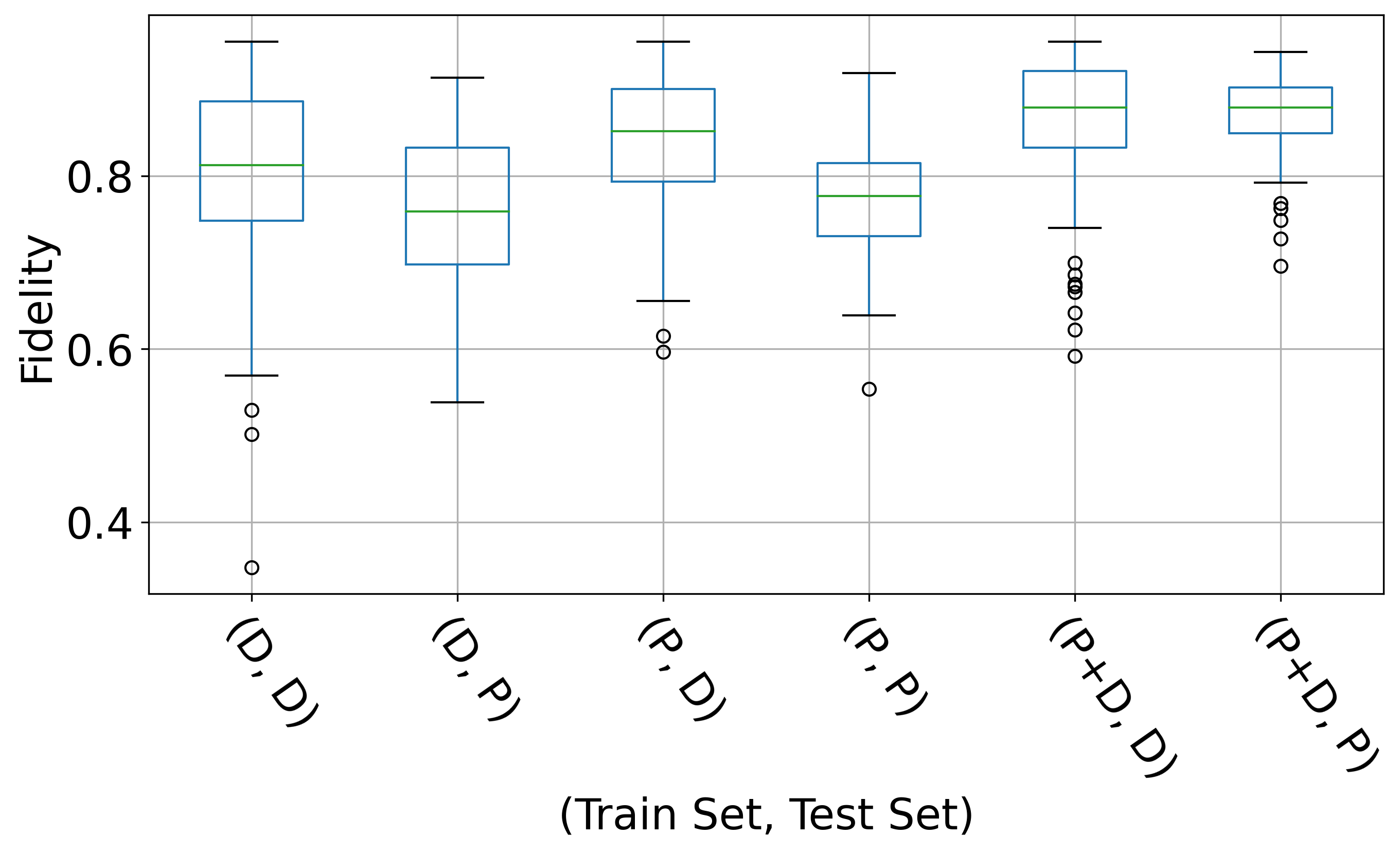}
        \caption{\small Fidelity (higher is better).}
        \label{fig:al_eval_fidelity}
    \end{subfigure}%
    \begin{subfigure}{.5\textwidth}
        \centering
        \includegraphics[width=\columnwidth]{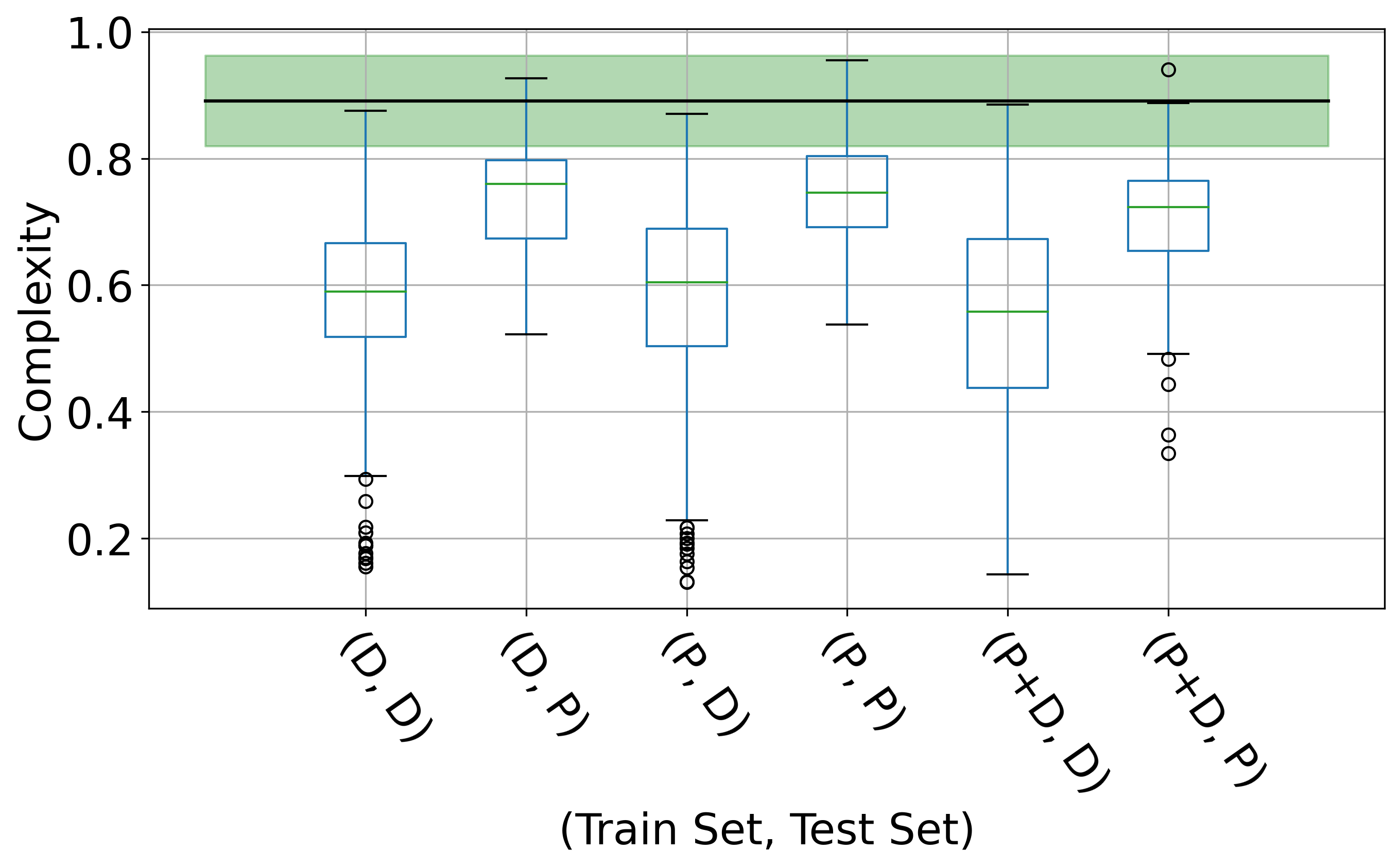}
        \caption{\small Complexity (lower is better).}
        \label{fig:al_eval_complexity}
    \end{subfigure}
    \caption{\small Figure~\ref{fig:al_eval_fidelity} shows the computed fidelity (coefficient of determination $R^2$ between the predictions by the global model $f$ and the local model $g$) scores for the evaluated explanations. Figure~\ref{fig:al_eval_complexity} shows the complexity (entropy of a distribution over the feature attribution weights) scores for the evaluated explanations. The green region shows the standard deviation of complexities for $1000$ random explanations, with the black line being the mean.
    }
    \label{fig:al_eval}
\end{figure*}
\begin{figure}
    \centering
    \includegraphics[width=\columnwidth]{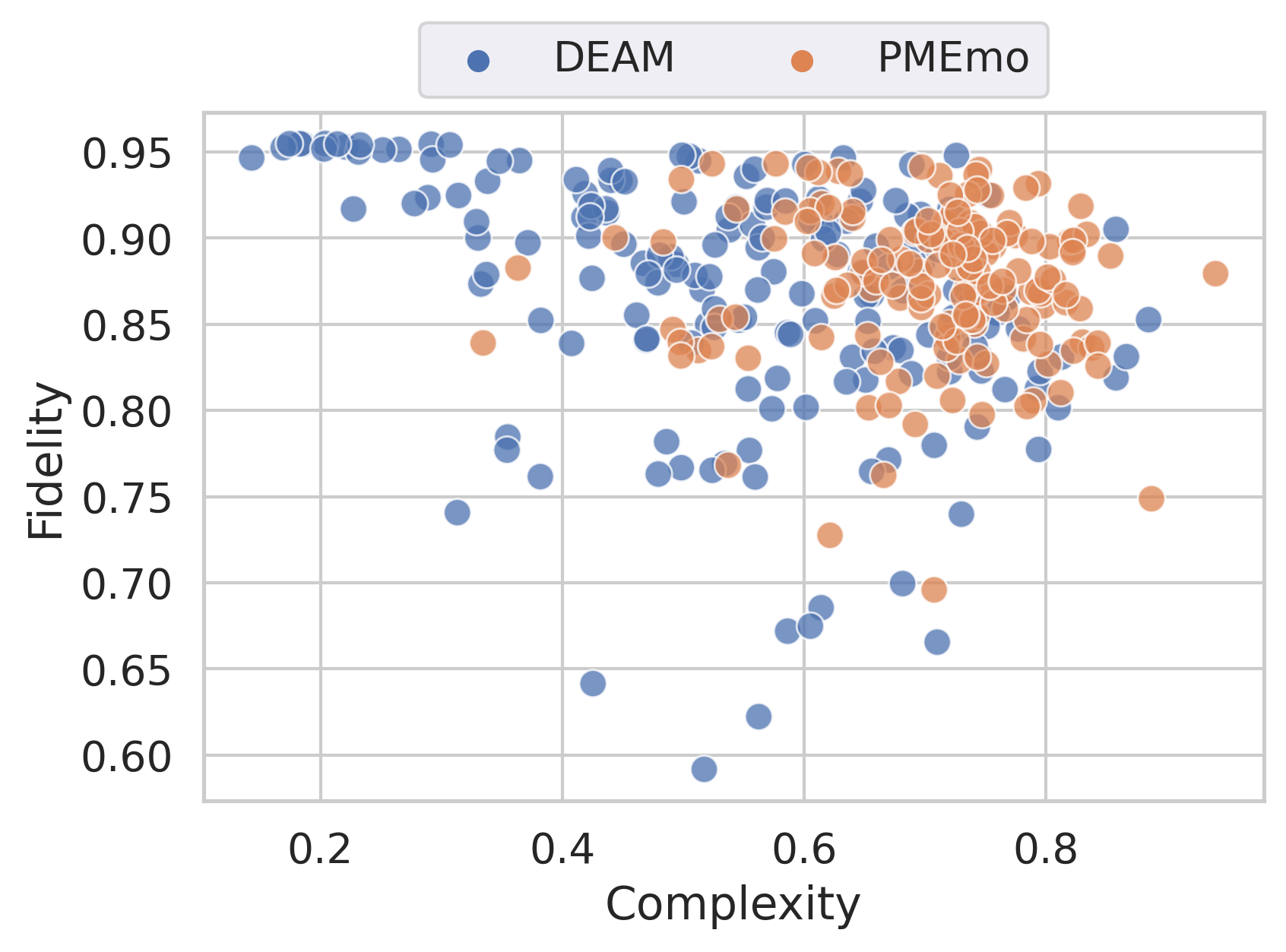}
    \caption{\small A more detailed view on the relationship between the fidelity score and complexity for the predictions of ``rhythmic stability'' for a model trained on both datasets. The color indicates the test set.}
    \label{fig:score_complexity}
\end{figure}

For the analysis, we compute predictions and explanations for all test examples and calculate the above mentioned metrics. The results for one mid-level feature (we picked ``rhythmic stability'' as it is used later on as an example) are summarized in Figures~\ref{fig:al_eval} and~\ref{fig:score_complexity}. We can see in Figure~\ref{fig:al_eval_fidelity} that the fidelity score (coefficient of determination) is relatively high across all combinations of models and test sets. The median score is $0.86$ across all explanations (including all mid-level features), the 25\%-quantile is at $0.78$. This means that for 50\% and 75\% of the explanations more than 86\%, and 78\%, respectively, of the variation in the dependent variable (mid-level prediction) can be predicted using the independent variables (instrument sources). 

Figure~\ref{fig:al_eval_complexity} shows the computed complexities, compared to a random baseline. Most explanations are far less complex than the random baseline.


The results shown in the previous figures suggested a relationship between the fidelity and complexity scores. Therefore we visualized the two metrics for all explanations computed for ``rhythmic stability'' for a model trained on the combined data sets in Figure~\ref{fig:score_complexity}. Although they seem related on a dataset level, the metrics do not look related when analyzed for each explanation separately, suggesting that indeed both are needed.


\begin{figure}
    \centering
    \includegraphics[width=\columnwidth]{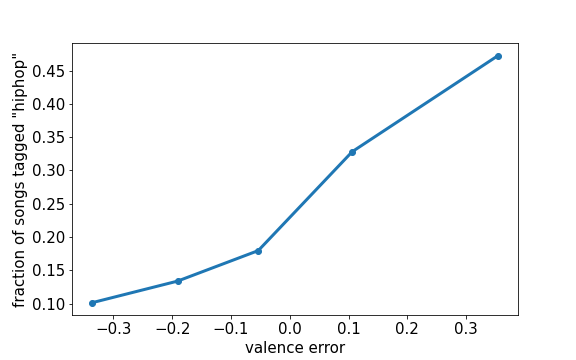}
    \caption{\small Fraction of hiphop songs in quantiles vs  the mean valence error of each quantile over PMEmo dataset (with model trained on DEAM)}
    \label{fig:quantiles}
\end{figure}

\section{Model Debugging}
\label{sec:model_debugging}
A practical use case of our explanation scheme is demonstrated in this section. We use the two-level explanations to understand why an improperly trained model might be overestimating the valence predictions for one particular genre.

\subsection{Setup}

First, we use a pre-trained tagger~\cite{musicnn} to predict genre tags for all the tracks in the three datasets mentioned in Section~\ref{sec:data}, since we do not have genre metadata for these datasets. This is only done in order to obtain an estimate of the genre-dependence of the emotion predictions later on. 
We then train two explainable models -- one on the DEAM dataset, and one on the combined DEAM and PMEmo dataset. The test set is a fixed but randomly chosen subset of the PMEmo dataset (with a mutually exclusive set of artists from the training set).
\begin{figure*}[h!]
    \begin{subfigure}{.5\textwidth}
        \centering    
        \includegraphics[width=\columnwidth]{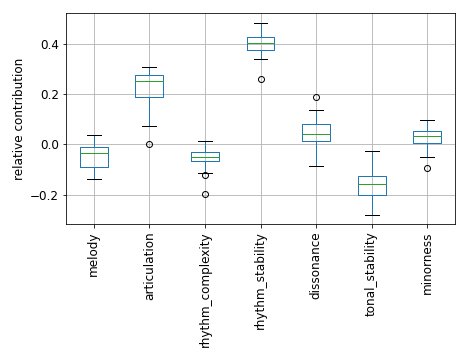}
        \caption{\small Trained on DEAM}
        \label{fig:rel_eff_biased}        
    \end{subfigure}%
    \begin{subfigure}{.5\textwidth}
        \centering
        \includegraphics[width=\columnwidth]{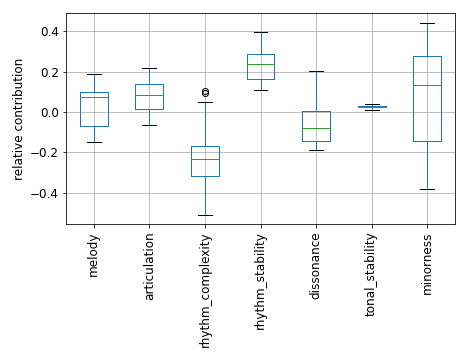}
        \caption{\small Trained on PMEmo+DEAM}
        \label{fig:rel_eff_combined}   
    \end{subfigure}
    \caption{\small Relative effects of the mid-level features for valence prediction for two models trained on different datasets, but tested on the same fixed subset of the PMEmo dataset.}
    \label{fig:rel_eff_both}
\end{figure*}

\subsection{Overestimated Valence for Hiphop}
When we take the model trained only on DEAM and use it to predict arousal and valence for the entire PMEmo dataset, we observe that the error in valence shows a pattern -- overestimations of valence primarily occur in hiphop songs, as shown in Figure~\ref{fig:quantiles}. 

\begin{figure}[h]
    \centering
    \includegraphics[width=\columnwidth]{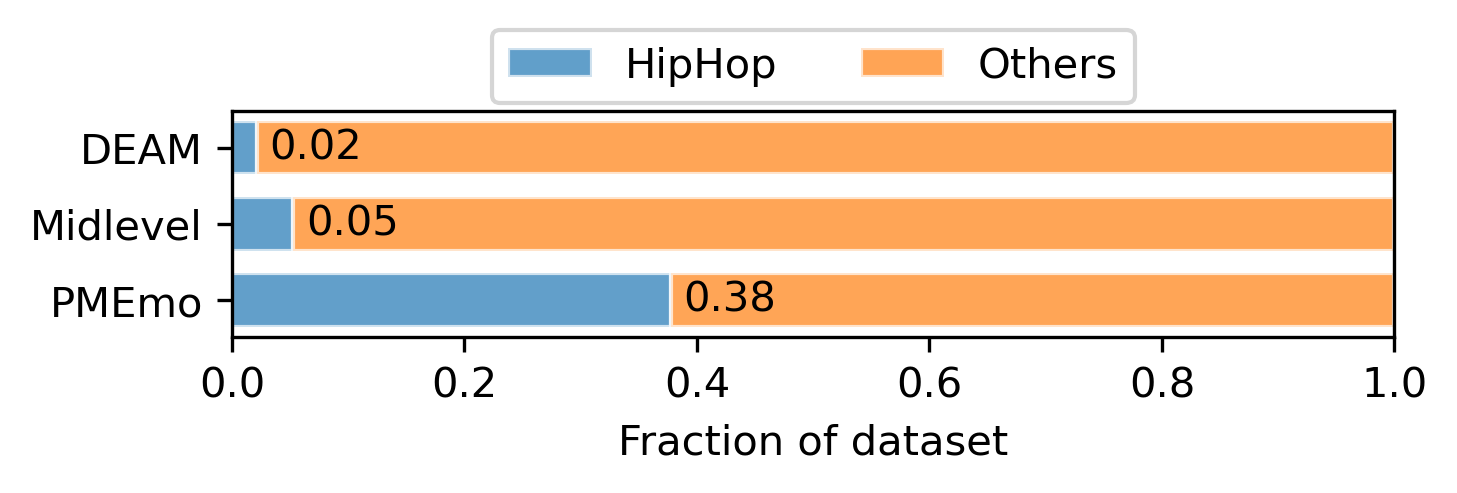}
    \caption{\small Compositions of datasets as fraction of songs tagged "hiphop" by a pre-trained auto-tagging model~\cite{musicnn}}
    \label{fig:genres}
\end{figure}

We can reason about relatively poor performance for hiphop songs based on the discrepancy between the training and testing sets in terms of genre composition. In Figure~\ref{fig:genres}, we can see that PMEmo has a large percentage of hiphop songs whereas both DEAM and Mid-level datasets have a small percentage. Since our model has not seen enough hiphop songs during training, it is to be expected that it does not perform well when it encounters hiphop during test. However, a question that is pertinent next is -- what is it about hiphop songs that makes our model overestimate their valence?

\subsection{Explaining Valence Overestimations Using Mid-level Features}
To answer this question, we first seek to understand which of the mid-level qualities can be attributed most to high valence predictions. This is the first level of our explanation system. We find these attributions by computing the effects of each mid-level feature on the valence predictions. The effect of a feature is simply the value of that feature multiplied by the weight of the linear connection between it and the target node. In our case, the target is valence and there are seven mid-level features that affect it. We are only interested in relative contribution of each feature, and so we divide each effect by the sum of the absolute values of the effects of all features and take the average across all test songs tagged ``hiphop''.

We observe that rhythmic stability has the maximum positive relative effect on the prediction of valence. Therefore, we select rhythmic stability for the next step of explanation.


\subsection{Explaining Rhythmic Stability Using Sources}
Once we have selected a mid-level feature as having the most positive relative effect on the valence, we would like to understand what musical constituents in the input can be attributed to positive contribution to that feature. To do this, we take the help of audioLIME and generate source based explanations for rhythmic stability. The sources available in the current implementation of audioLIME\footnote{\url{https://github.com/CPJKU/audioLIME}} are vocals, drums, bass, piano, and other. 

We find that vocals are a major contributing source for the rhythmic stability predictions for the hiphop songs. For songs tagged as other genres, contributing sources are more distributed. 

\subsection{Re-training the Model with Target Data}

Bringing together our two types of
explanations, we can reason that the high valence predictions for hiphop songs is due to overestimation of rhythmic stability, which, in this case, can be attributed to the vocals. While there is a lot of diversity in the style of \textit{rapping} (the form of vocal delivery predominant in hiphop), it has been noted that rappers typically use stressed syllables and vocal onsets to match the vocals with the underlying rhythmic pulse~\cite{ohriner2019lyric, adams2009metrical}. These rhythmic characteristics of vocal delivery (that constitutes ``flow'', and may add metrical layers on top of the beat) contribute strongly to the rhythmic feel of a song. 
The positive or negative emotion of hiphop songs is mostly contained in the lyrics -- the style of vocal performance does not necessarily express or correlate with this aspect of emotion. Therefore, it makes sense that a model which has seen few examples of hiphop during training should wrongly associate the prominent rhythmic vocals of hiphop to high rhythmic stability and in turn high valence. A model that has been trained with hiphop songs included, we expect, would place less importance on rhythmic stability for the prediction of valence, even if the vocals might still contribute significantly to rhythmic stability. Thus, we expect the relative effect of rhythmic stability for valence to decrease in such a model. 

This is exactly what we observe on a model trained with the combined PMEmo+DEAM dataset. The average relative effects are shown in Figure~\ref{fig:rel_eff_combined} and we can see that the relative effect of rhythmic stability has decreased while those of minorness, melody, and tonal stability have increased. Thus, the model changed in a way that was in line with what we expected from the analysis of our two-level explanation method.

\begin{figure}[t]
    \centering
    \includegraphics[width=\columnwidth]{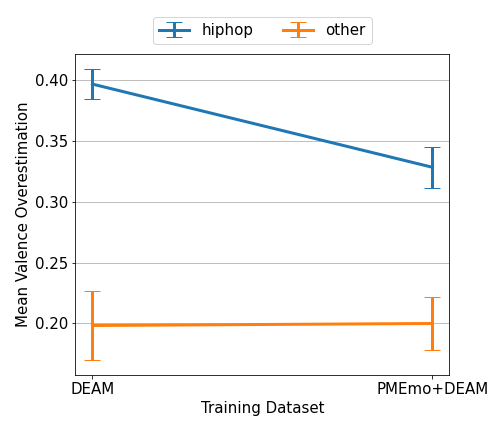}
    \caption{\small Mean valence overestimations for two models trained on different datasets, but tested on the same fixed subset of the PMEmo dataset.}
    \label{fig:mean_overestimations}
\end{figure}

Looking at mean overestimations (Figure~\ref{fig:mean_overestimations}) in valence for hiphop and other genres for models trained on DEAM and PMEmo+DEAM shows that valence overestimations of hiphop songs have decreased substantially, without negatively affecting the predictions on other genres\footnote{Code for reproducing model debugging experiments is available at \url{https://github.com/shreyanc/model_debugging}}.


\section{Conclusion and Future Work}
In this paper, we proposed a method to explain music emotion models in an intuitive way using components from low- and mid- levels of the hierarchy of musical concepts by combining audioLIME, which uses input audio sources as explanatory components, with intermediate layer based explanations. We also demonstrated its potential as a tool for model debugging and explaining model behaviour.

This points us towards exploring this method further and getting more granular explanations 
as a way of improving the effectiveness of this system for MIR. An immediate next step that we are currently pursuing is to extend audioLIME to provide explanations in the form of temporal segments using semantic music segmentation, along with the sound sources. 

We are also looking at explaining emotion conveyed in classical piano performances, which pose particular challenges -- including the non-availability of training data, where transfer learning of explanatory features becomes necessary~\cite{chowdhury2021towards}.


\vspace{4mm}
\begin{acknowledgments}
	This work is supported by the European Research Council (ERC) under the
EU’s Horizon 2020 research \& innovation programme under grant
agreement No. 670035 (“Con Espressione”), and the Federal State of Upper Austria
(LIT AI Lab). 
\end{acknowledgments}


\bibliography{smc2021bib}
\end{document}